\documentclass[aps,preprint,epsfig]{revtex4}

\usepackage{graphicx}
\usepackage{dcolumn}
\usepackage{bm}

\newcommand{\bq}{\begin{equation}}
\newcommand{\eq}{\end{equation}}
\newcommand{\bqa}{\begin{eqnarray}}
\newcommand{\eqa}{\end{eqnarray}}
\newcommand{\nn}{\nonumber \\}

\begin{document}
\draft
\title{
A dual gravity study of the 2+1D compact U(1) gauge theory 
coupled with strongly interacting matter fields
}

\author{Sung-Sik Lee and Xiao-Gang Wen}
\address{Department of Physics, Massachusetts Institute of Technology,\\
Cambridge, Massachusetts 02139, U.S.A.\\}
\date{\today}
       
\begin{abstract}
We consider the D2-brane probe action in the gravity background dual to N
coincident Dp-branes by treating the separation between the D2- and Dp-branes
as a nondynamical parameter for $p=2,4,6$.  
The gauge coupling, the core size of a non-BPS instanton
and the mass gap of the compact U(1) gauge theory in the D2-brane are
determined as a function of the separation in the type IIA gravity region.  
The results are interpreted in terms of the 2+1D U(1) gauge theory coupled with
the matter fields which are also strongly coupled with the p+1D SU(N) gauge field.
It is shown that strong coupling of the matter fields to the SU(N) gauge field
can drastically modify their screening of the U(1) gauge field.
The non-perturbative dependence of the U(1) gauge coupling on the energy scale
is obtained.
\end{abstract}
\maketitle

\newpage

\section{Introduction} 
Polyakov have shown that there is no deconfinement phase for the pure
2+1D compact U(1) gauge theory\cite{POLYAKOV77}.  In the
confinement phase instantons proliferate and the gauge field acquires a mass
gap.  After the seminal work\cite{POLYAKOV77} a good deal of theoretical
efforts have been devoted to the question of how the presence of matter field
with fundamental charge modifies the dynamics of the U(1) gauge
field\cite{EINHORN,FRADKIN79,NAGAOSA93,MUDRY,NAYAK,NAGAOSA00,ICHINOSE,WEN02,RWspin,HERBUT,KLEINERT,HERMELE}.
The dynamics of the U(1) gauge field crucially depends on the number and the
dynamics of the matter fields\cite{IOFFE,MURTHY,WEN02,RWspin,SENTHIL04,HERMELE}.
Theoretical analysis is most feasible if there are a large number of matter
fields.  One loop calculations show that the gauge coupling is renormalized to
be $g^2 \sim \frac{\Lambda}{N}$ with $N$, the number of matter fields and
$\Lambda$, the mass of the matter fields\cite{MURTHY}.  Consequently the
instanton acquires a large scaling dimension ($\sim N$) and becomes irrelevant
at the critical point in the limit $\Lambda \to 0$
\cite{IOFFE,WEN02,RWspin,SENTHIL04,HERMELE}.  
Then it is interesting to ask how a change in the dynamics of matter field
affects the dynamics of the U(1) gauge field.  
The self-interaction of massive matter fields was shown to qualitatively modify the short distance
potential between test charge in the non-compact 2+1D quantum
electrodynamics\cite{GHOSH,ABREU}.  An alternative way of modifying the
dynamics of matter fields is to put the matter fields under a strong
additional gauge interaction.
In this paper, we are going to consider a system of 2+1D $U(1)$ gauge
theory coupled with matter fields in 2+1D where the matter fields in turn
interact strongly with a $SU(N)$ gauge field in p+1D.
(Here the 2+1D space-time is a subspace of the p+1D space-time with $p=2, 4, 6$.)
When $p = 4, 6$ ($p = 2$) the SU(N) gauge coupling becomes weak at low (high) energy.
In this regime the theory reduces to the aforementioned
2+1D U(1) gauge theory coupled with matter fields.
Then how will the dynamics of the 2+1D U(1) gauge field be modified 
at high (low) energy for $p = 4, 6$ ($p = 2$) where the SU(N) gauge coupling becomes strong ?
Usual perturbative picture is not suitable to describe the strong coupling effect. 
The aim of the present paper is to examine the non-perturbative effect of
the strong SU(N) gauge coupling on the 2+1D U(1) gauge field.

For some strongly coupled gauge theories, including the one under
consideration, it is advantageous to use dual string
theory\cite{AHARONY}.  The exact duality between gauge and string theories has
been anticipated from the observation that the Wilson loop in gauge theory
satisfies a loop equation of string\cite{POLYAKOV98}.  
The first concrete example for
this idea was conjectured as a duality between the type IIB string theory in
the anti-de Sitter space and ${\cal N}=4$ supersymmetric SU(N) gauge theory in
3+1D\cite{MALDACENA,GUBSER,WITTEN}.  The duality has opened a
variety of possibilities for a new understanding on many strong coupling
phenomena of gauge theories\cite{AHARONY}.  From the dual gravity description
the confining nature of the 2+1D SU(N) gauge theory has been
confirmed\cite{WITTEN2}.  Recently the idea has been applied to construct
QCD-like gauge theory including fundamental matter
fields\cite{KARCH,KRUCZENSKI,BABINGTON,CHERKIS,NUNEZ,ERDMENGER}.
Most recently dual gravity backgrounds have been found for an infinite family of quiver gauge theories\cite{BENVENUTI}.

The field theory of our interest is a nonsupersymmetric theory.  
It contains a p+1D SU(N) gauge theory with matter fields in
the adjoint representation of the SU(N) gauge group,
and a U(1) gauge theory that lives on a 2+1D subspace.
It also contain matter fields on the 2+1D subspace that carry fundamental
charges for both U(1) and SU(N) gauge fields.  To understand the dynamics of
the $U(1)$ gauge field, we would like to integrate out the SU(N) gauge field
and the matter fields to obtain an effective theory of the U(1) gauge field.
However, this is not easy to do in the strong coupling limit.  In this paper, we
like to show that, in the large $N$ limit, we can obtain the effective action
using a duality relation between the above field theory and
D-brane in superstring theory.

The above 2+1D/p+1D U(1)/SU(N) gauge theory 
has a dual description in terms of superstring theory where we consider a probe
D2-brane lying parallel to a large number of Dp-branes in type IIA superstring
theory.  However the full field theory describing the brane system is larger
than the field theory of our interest.  Fortunately, it is possible to study a
reduced field theory from the brane configuration in the probe limit, as will
be explained below.  We first identify the full degrees of freedom in the field
theory for the brane configuration, then explain how we obtain the reduced
field theory of our interest.

The low energy field theory on the D2-brane is the 2+1D U(1) gauge theory.  The
U(1) gauge field comes from open string with its two ends on the D2-brane.  The
open strings connecting different Dp-branes give rise to p+1D SU(N) gauge
fields on the Dp-brane.  The matter fields, carrying fundamental charges for
both U(1) and SU(N) gauge fields, come from the open strings that connect the
D2- and Dp-branes.  There are also U(1)/SU(N) neutral scalars coming from the
open strings with its two ends on the D2-branes.  They describe fluctuations in
the relative position of the D2 and Dp-branes.  In the supersymmetric case
($p=2,6$) there are also fermionic partners to all of the bosonic modes.  These
are the degrees of freedom of the full field theory for the brane
configuration.

We consider the probe action of the D2-brane in the gravity background 
dual to the $N$ Dp-branes.
In the probe limit, the back reaction of the D2-brane to metric is not included.
More specifically, the fluctuations of the neutral scalars are frozen
by fixing the position and the flat shape of the probe brane.  We treat the
separation between the D2 and Dp branes as a non-dynamical parameter ignoring
the fluctuations.  The fermionic modes on the D2-brane do not have geometrical
meaning like the position of brane because they can not have vacuum expectation
values.  Thus we just ignore the fluctuations of those modes in the effective
action.  Certainly, we also ignore the fluctuations of 2+1D U(1) gauge field.

We see that in the probe limit, the probe action only includes the effect of
integrating out all the p+1D fields including the SU(N) gauge field,
 and the fundamental matter fields,
but not the fluctuations of the 2+1D U(1) gauge field, neutral scalars, and
their fermionic partners on the D2-brane.  Thus one can regard the probe action
with background U(1) gauge field as an effective action for the U(1) gauge
field which is obtained by integrating out the SU(N) gauge field 
along with other p+1D fields and the fundamental matter field.
This, in turn, can be interpreted as the effective action obtained 
from the reduced field theory which includes all the degrees of freedom
of the full field theory except for the neutral 
scalars and fermions coming from the strings
with their two ends attached to the D2-brane.

In this approach, the effective coupling strength of the SU(N) gauge
interaction and the mass of the matter fields can be tuned independently by the
separation between the branes and the string coupling constant.  Using the
resulting 2+1D U(1) effective action, we can examine how the U(1) gauge
coupling and the mass gap of the U(1) gauge theory change as the energy scale
(set by the separation between branes) varies.  The mass gap of the U(1) gauge
theory is generated by the proliferation of the U(1) instanton.

It should be emphasized that the U(1) effective action is not meant to describe
the full field theory of the branes.  We use the brane configuration as a tool
to integrate out the strongly coupled matter fields and the SU(N) gauge field.
In the full field theory of the brane, the neutral scalars and the fermionic
modes should be allowed to fluctuate along with the U(1) gauge field.  This
makes significant differences in the dynamics of instanton.  First, in the full
field theory the neutral scalars acquire space-time dependent expectation value
in the presence of U(1) instanton.  In the brane picture, the probe brane is
bent near the instanton, which, in turn, modify the interaction between
instantons.  Second, the presence of the fermionic zero modes on the D2-brane
associated with the underlying supersymmetry for $p=2, 6$ will suppress the
multi-instanton effects in the full field theory.  As a result, the U(1) photon
(equivalently, the scalar dual to the photon in 2+1D) remains massless with
$16$ sypercharges\cite{SEIBERG}.  However there are multi-instanton effects in
the reduced non-supersymmetric field theory of our interest. This is because
there are no such neutral scalars and fermions in our field theory model.

There are unbroken supersymmetries in the D2/Dp system for $p=2$ and $6$.  
The full field theory includes not only the (p+1)D super SU(N) gauge theory
but also the 2+1D super U(1) gauge theory.
With the suppression of fluctuations of neutral scalars and the U(1) gaugino on the D2-brane,
the reduced field theory becomes a non-supersymmetric 2+1D U(1) gauge theory.
However the U(1) gauge theory is still coupled to the supersymmetric (p+1)D SU(N) gauge theory with the fundamental matter fields.
We will examine how the dynamics of the U(1) gauge field is affected by the fundamental matters which are strongly coupled to the super SU(N) gauge field.

For $p=4$ there is no supersymmetry.  The absence of supersymmetry makes the
D2/D4 system unstable.  Eventually the D2-brane will collapse to the D4-branes
and will be dissolved into flux in the D4-branes.  In the full unstable field
theory there are tachyons describing the transverse fluctuations of the
D2-brane.  For the purpose of exploring the full non-perturbative structure of
string theory it is essential to consider the tachyon condensation\cite{ASEN}.
Here we freeze the unstable fluctuations and study the field theory describing
the fluctuations along the stable direction.  Conceptually this is
similar to the Gliozzi-Scherk-Olive(GSO) projection in constructing
tachyon-free string theories out of full string spectrum including tachyons.
However the identification of stable field theory is less clear in the present
case because the tachyonic modes are coupled to other stable modes while there
is no such coupling in the GSO projection.  Thus it is not clear {\it in
priori} whether the D2/D4 brane with fixed distance describes the 2+1D/4+1D
U(1)/SU(N) gauge theory.  In this paper we present a clue that it may be really
the case.

\section{Effective action of the U(1) gauge theory from dual gravity approach}

First we consider a general configuration involving D2 and Dp branes in
10-dimensional type IIA string theory with $p$ an even integer.  
Then we will discuss $p=2, 6, 4$ cases in the order that the number of supercharges is lowered.

Consider N coincident Dp-branes and one D2-brane where the Dp-branes are
extended in $0,1,...,p$ directions and the D2-brane, in $0,1,2$ directions.
In the field theory limit\cite{MALDACENA,ITZHAKI} the low energy theory
consists of two decoupled theories : 1) 9+1D gravitational
theory, and 2) 2+1D U(1) gauge theory on the D2 brane and
p+1D SU(N) gauge theory on the Dp branes.  The U(1) and SU(N)
gauge theories are coupled with each other through matter fields.  The matter
fields come from stretching strings between the D2-brane and the Dp-branes,
and thus they are extended only in the 0,1,2 directions.  They carry
fundamental charges for both U(1) and SU(N) gauge fields.  Since there are $N$
different possibilities of the string's ending on the Dp-branes, the number of
matter fields is proportional to $N$.  For each one of $N$ there are a few
light string modes and a tower of infinitely massive modes.

We replace the N Dp-branes with a gravitational background.  The Euclidean
metric for the N Dp-branes which are located at $x^{p+1} = x^{p+2} = ... = x^9
= 0$ is (in string frame)\cite{AHARONY}
\bq
ds^2 = 
\frac{1}{\sqrt{H_p(r)}} \left[ \sum_{i=0}^{p} (dx^i)^2 \right] 
+ \sqrt{H_p(r)} \left[ \sum_{j=p+1}^{9} (dx^j)^2 \right], 
\label{metric}
\eq   
where $H_p(r) = 1 + \left( \frac{L_p}{r} \right)^{7-p}$ with $r = \sqrt{
\sum_{j=p+1}^{9} (x^j)^2 }$ and $L_p^{7-p} = \frac{ (2\pi)^{7-p} g_s l_s^{7-p}
N}{ (7-p) \Omega_{8-p} }$.  $g_s$ is the string coupling in the asymptotic
region ($r \rightarrow \infty$), $l_s$, the string length scale and
$\Omega_{d}$, the volume of the unit d-dimensional sphere.  The classical
gravity approximation is reliable if the curvature is small in the unit of
string length and the local string coupling is small,
\bq
| {\cal R} | l_s^2  <<  1,  \mbox{\hspace{0.5cm}} e^\phi  <<  1,
\label{smallRg}
\eq
where ${\cal R}$ is the scalar curvature of the metric (\ref{metric}) and
$e^\phi = g_s H_p^{\frac{(3-p)}{4}}$ with $\phi$, the dilaton field.

In the probe limit ($N >> 1$) the back reaction of the D2-brane on the metric
is negligible.  The effective action of the U(1) gauge field on the D2-brane
is given by the Dirac-Born-Infeld action, 
\bqa
\Gamma & = & \frac{1}{(2\pi)^2 l_s^3}  \int dx^3 e^{-\phi}
\sqrt{det(G^{ind}_{\mu \nu} + 2\pi l_s^2 F_{\mu\nu})},
\label{Gamma1}        
\eqa
where $G^{ind}_{\mu \nu}$ is the induced metric on the probe D2-brane.
Here we suppress the transverse fluctuations of the D2-brane and treat $r$ as a parameter.  
We take the field theory limit\cite{MALDACENA,ITZHAKI},
\bqa
l_s & \rightarrow & 0, \nn 
g_{YM}^2 & = & (2 \pi)^{p-2} g_s l_s^{p-3} = {\mbox {\rm fixed}}, \nn 
\Lambda & = & \frac{r}{l_s^2} = {\mbox {\rm fixed}}.
\label{limit}
\eqa 
Here $g_{YM}^2$ is the coupling constant for the p+1D SU(N) gauge
theory on the Dp-branes.  $\Lambda$ is the mass scale associated with the
tension of string stretching between the D2 and Dp-branes.  

The resulting field theory contains an $SU(N)$ gauge field in $p+1$
dimensional space-time and a $U(1)$ gauge field in $2+1$ dimensional space-time
which is a subspace of the $p+1$ dimensional space-time.  The field theory
also contain bosonic and fermionic matter fields in $2+1$ dimension 
that carry both the SU(N) gauge charge and the U(1) gauge charge.
The mass scale of the matter field and the high energy cut-off scale of the field theory is of order $\Lambda$.
In this paper, we like to understand the dynamics of the U(1) gauge field in $2+1$ dimensions.

The Dp/Dq ($p>q$) brane systems have been considered in order to add fundamental matters to the q+1D gauge theories\cite{KARCH,KRUCZENSKI,BABINGTON,CHERKIS,NUNEZ,ERDMENGER}.
In the previous studies\cite{KARCH,KRUCZENSKI,BABINGTON,CHERKIS,NUNEZ,ERDMENGER} the gauge coupling in the light (Dq) brane was taken to be finite in the field theory limit.
In this limit the gauge coupling in the heavy (Dp) brane vanishes.
The gauge symmetry in the heavy brane becomes a global flavor symmetry.
In our case we do the opposite in order to study the effect of strong coupling in the heavy brane.
We take the gauge coupling in the heavy brane finite. 
Then the gauge coupling in the light brane becomes infinite. 
It is noted that the bare 2+1D gauge coupling is $g_s l_s^{-1} \rightarrow \infty$ in the field theory limit (\ref{limit}) for $p=4$ and $6$.
Then how do we obtain a finite 2+1D gauge coupling ?
This is possible because the fundamental matter fields renormalize the gauge coupling to a finite value. 
In other words there is no bare kinetic energy term for the 2+1D gauge field but it is generated by the fluctuations of the fundamental matter fields.

In the strong coupling limit of the SU(N), we take the 't Hooft limit where
the effective Yang-Mills coupling $g_{eff}^2 = g_{YM}^2 N \Lambda^{p-3}$ is
fixed in the large N limit\cite{THOOFT}.
We like to obtain
the low energy effective action for the U(1) gauge field 
in this limit by integrating out the SU(N) gauge fields and the
matter fields.
Instead of directly integrating out the SU(N) gauge fields and the
matter fields, we go back to the string theory and integrate out all the
string modes to obtain the
the low energy effective action for the U(1) gauge field:
\bqa
\Gamma & = & 
\frac{M^2}{g^2} \int dx^3 \sqrt{ F^2 + M^4 }, 
\label{Gamma}        
\eqa
where
\bqa
g^2  & = &  [ (7-p) \Omega_{8-p} ]^{\frac{(p-2)}{4}}
(2\pi)^{\frac{(p-2)(2p-13)}{4}} g_{YM}^{\frac{(6-p)}{2}} N^{\frac{(2-p)}{4}}
\Lambda^{\frac{(7-p)(p-2)}{4}}, \nn 
M^4  & = &  (2 \pi)^{2p-11} (7-p)
\Omega_{8-p} N^{-1} g_{YM}^{-2} \Lambda^{7-p}, 
\label{obs}
\eqa 
and $F^2 \equiv \sum_{\mu > \nu} F_{\mu \nu} F_{\mu \nu}$, the square of the U(1) gauge field strength.
$\Gamma$ corresponds to the effective action generated by vacuum fluctuations of strings in the background of the U(1) gauge field on the probe D2-brane and the gravitational field dual to the Dp-branes\cite{DBI}. 
In the weak string coupling limit ($e^{\phi} <<1$) the leading contributions come from the disk diagrams of string world sheet.
In the field theory side the weak string coupling limit corresponds to the 't Hooft limit, and the disk diagrams to the planar diagrams (see Fig. 1(a)).
Higher order diagrams (e.g., cylinder diagrams) in string theory corresponds to non-planar diagrams in field theory (see Fig. 1(b)).
The disk diagram is order of $e^{-\phi} \sim N$ and the cylinder diagram, $e^{0} \sim 1$. 
Thus the Dirac-Born-Infeld action which is order of $e^{-\phi}$ captures the fluctuations of matter fields and the SU(N) gauge fields in the leading order of $N$ in the field theory side\cite{AHARONY}.
The effects of matter fields and the SU(N) gauge field are encoded in the nontrivial metric background of Eq.(\ref{Gamma1}) and in $g^2$ and $M$ of Eq.(\ref{Gamma}).  
Note that it is possible to regard the brane position as a non-dynamical parameter 
because the fluctuations of strings with their two ends on the D2-brane are negligible
 by factor of $1/N$ in the probe limit.
The derivative terms such as $(\partial F)^n$ are ignored in this action.  
If the action is expanded in $F^2$, the coefficient of the quadratic term becomes $\frac{1}{2 g^2}$.  
Thus $g^2$ is identified as the gauge coupling of the 2+1D U(1) gauge theory.  
$M$ is the mass scale above which higher order terms become important.  
$M$ can be also identified as the size of non BPS instanton as will be discussed later.
The conditions for the small curvature and small string coupling
(\ref{smallRg}) become\cite{ITZHAKI}
\bq
1 <<  g_{eff}^2 << N^{\frac{4}{7-p}}.
\label{the_range0}
\eq

Now we determine the mass gap of the 2+1D compact U(1) gauge
field as a function of $\Lambda$.  
We have to consider instantons because of the compactness of the gauge field.  
It is emphasized again that the instanton we consider here is non-supersymmetric even though the background is supersymmetric for $p=2, 6$.
This is because the excitations of scalar fields are suppressed.
The instanton is an event localized in space-time where the
U(1) flux changes by $2\pi$\cite{POLYAKOV77}.  Using the dual field strength
$b_\mu = \frac{1}{2} \epsilon_{\mu \nu \lambda} F_{\nu \lambda}$, we divide
$b$ into the longitudinal part and the transverse part,
\bq
b_\mu = b_\mu^{in} + (\partial \times a)_\mu,
\eq
where the longitudinal part $b_\mu^{in}$ is contributed from the instantons
and satisfies
\bq
\partial \cdot b^{in} = 2 \pi \rho,
\label{inst}
\eq
with $\rho$, the instanton density.  We consider one instanton of charge $q$
at the origin with $q$, an integer.  The instanton action is obtained by
minimizing the effective action Eq.(\ref{Gamma}) with respect to the
transverse field $a$.  The resulting equation of motion for $a$,
\bq
\partial \times \frac{ b^{in} + \partial \times a }{\sqrt{ (b^{in} + \partial \times a)^2 + M^4 }} = 0
\eq
is solved by introducing a dual scalar field $\xi$,
\bq
\frac{ b^{in} + \partial \times a }{\sqrt{ (b^{in} + \partial \times a)^2 + M^4 }} = \partial \xi.
\eq
From Eq.(\ref{inst}) $\xi$ satisfies
\bq
\partial \cdot \left(    
\frac{ \partial \xi }{ \sqrt{ 1 - (\partial \xi)^2 } } 
\right)
= 2 \pi q \frac{\delta^{(3)}(x)}{M^2},
\label{flux}
\eq
with $\delta^{(3)}(x)$, the three-dimensional delta function resulting in
\bq
(\partial_r \xi)^2 = \frac{1}{ (\sqrt{2/q} M r)^4 + 1 }.
\eq
For $r << M^{-1}$ the dual scalar field increases linearly with distance.  
On the other hand for $r >> M^{-1}$ we obtain $\xi \sim 1/r$.  
Thus we identify the length scale $M^{-1}$ as the core size of instanton.  
From Eq.(\ref{Gamma}) the
instanton action is readily obtained to be
\bqa
I_{c}(q) & = &
\frac{M^4}{g^2} \int d^3 x \left[ \frac{1}{\sqrt{ 1 - (\partial \xi)^2 } } - 1   \right] \nn
& = &  \frac{2 \pi M q^{\frac{3}{2}}}{g^2}  \int_{0}^\infty dy [ \sqrt{ 4y^4 + 1} - 2y^2 ] 
\approx  5.5 \frac{q^{\frac{3}{2}} M}{g^2}.
\label{core}
\eqa
It is noted that the action of instanton is finite without short distance
divergence and that the action is proportional to the charge $q$ with a fractional power $\frac{3}{2}$.  
Both of these features are due to the higher
order terms of field strength in the effective action (\ref{Gamma}) which
become important near the center of the instanton.
It is noted that the energy scale associated with the instanton core is smaller than the cut-off scale, that is, $M \sim \frac{\Lambda}{ g_{eff}^{1/2} } << \Lambda$.
Thus the core structure of the instanton can be reliably studied from the effective action (\ref{Gamma}) as far as $\Lambda^{-1} << r$.

Now we consider many instantons.  Eq.(\ref{flux}) is modified as
\bq
\partial \cdot \left(    
\frac{ \partial \xi }{ \sqrt{ 1 - (\partial \xi)^2 } } 
\right)
= \frac{ 2 \pi}{M^2}  \sum_a q_a \delta^{(3)}(x-x_a).
\label{flux2}
\eq
If the distance between instantons is much larger than $M^{-1}$ the dual
scalar field becomes
\bq
\xi(x) \approx \frac{1}{2M^2} \sum_a \frac{q_a}{|x-x_a|}
\eq
leading to the Coulomb interaction between instantons\cite{POLYAKOV77},
\bq
\Gamma  =  \sum_a I_c(q_a) + \frac{\pi}{g^2} \sum_{a>b} \frac{q_a q_b}{|x_a - x_b|}.
\eq
Owing to the screening property of the 3D Coulomb gas the
$\frac{1}{x}$ potential is screened to be $\frac{e^{-m_c x}}{x}$ where $m_c$,
the mass gap of the U(1) gauge field\cite{POLYAKOV77}.  With $M^{-1}$
identified as a cut-off length scale for instanton the mass gap is given by
$m_c^2 \sim \frac{M^3}{g^2} e^{-I_c}$ with $I_c$, the instanton action with
unit charge\cite{POLYAKOV77,GOPFERT}.  
Using Eq.(\ref{obs}) the mass gap is obtained to be
\bq
m_c^2  \sim g_{YM}^{\frac{4}{3-p}} \lambda^{\frac{(p-5)(p-7)}{4}}  e^{-c_p \lambda^{\frac{(p-3)(p-7)}{4}} },
\label{mass}
\eq
where 
$\lambda = \Lambda (g_{YM}^2)^\frac{1}{(p-3)} N^{\frac{1}{p-7}}$ and 
$c_p = (2 \pi)^{\frac{(2p+3)(11-p)}{4}} [ (7-p) \Omega_{8-p} ]^{\frac{3-p}{4}}  \int_0^\infty dy [ \sqrt{ 1+ 4y^4} - 2 y^2 ]$.

\subsection{$p=2$} 
The full field theory is the SU(N+1) super Yang-Mills theory with $16$ supercharges where the gauge group is broken to $SU(N) \times U(1)$ for nonzero $\Lambda$.
The vector multiplet consists of the gauge field, 7 scalars and 8 Majorana spinors.
As discussed in the introduction we suppress the fluctuations of the scalars
and fermions in the U(1) sector of the vector multiplet in order to study
non-supersymmetric U(1) gauge theory.  The resulting theory is a 2+1D field
theory with a $U(1)$ gauge field, a $SU(N)$ gauge field, and some
bosonic/fermionic matter fields in the fundamental representation of
$U(1)\times SU(N)$ and adjoint representation of $SU(N)$.  The DBI action
(\ref{Gamma}) is the effective action for the U(1) gauge boson on the probe
D2-brane after the supermultiplets on the N D2-branes and the stretched string
modes are integrated out.  This corresponds to integrating out the $SU(N)$ gauge
field and the matter fields in the field theory.
The mass of the stretched string modes is given by $\Lambda$. 
This configuration is stable because the gravitational attraction is balanced by the coupling to the Ramond-Ramond field which we did not show in (\ref{Gamma}).
The conditions for the small curvature and small string coupling (\ref{smallRg})
becomes 
\bq
g_{YM}^2 N^{\frac{1}{5}} <<  \Lambda  <<  g_{YM}^2 N.
\label{the_range2}
\eq
For $\Lambda > g_{YM}^2 N$ the curvature becomes large in string unit and gravity solution is not reliable.
Instead perturbative field theory is reliable in this UV limit.
For $\Lambda < g_{YM}^2 N^{\frac{1}{5}}$ the local string coupling becomes large and the 11-th dimension of the M-theory appears\cite{ITZHAKI}.
We will concentrate only on the IIA gravity description in the range (\ref{the_range2}).
The U(1) gauge coupling and the inverse size of the instanton is given by
\bq
g^2 = g_{YM}^2 , \mbox{\hspace{0.5cm}} M = \left( \frac{\Lambda^5}{24 \pi^4 g_{YM}^2 N }  \right)^{1/4}.
\label{obs2}
\eq
The original Yang-Mills coupling $g_{YM}^2$ is restored for the U(1) sector of the SU(N+1) gauge theory 
as expected.
There is no loop correction to the U(1) gauge coupling.
This is because the integrated SU(N) gauge field and the matter field have $16$ supercharges.
The flow of the U(1) gauge coupling is solely determined by the dimensional scaling as is shown in Fig. 2(a).
The loop correction is absent also in the regime of the perturbative SU(N) gauge theory.
Thus the dimensionless U(1) gauge coupling is likely to behave as $\Lambda^{-1}$ 
in the whole range of the energy scale including both the weak and strong (IIA gravity) coupling regimes (see Fig. 2(a)).

One can readily obtain the action of the instanton and the mass gap of the U(1) gauge field from (\ref{mass}).
The U(1) instanton considered here is different from the supersymmetric instanton of the full SU(N+1) gauge theory\cite{INSTANTON}.
We are considering a non-supersymmetric instanton which involves the excitation of only the U(1) gauge field on the probe brane.
The supersymmetric instanton\cite{INSTANTON} corresponds to Euclidean D0-brane stretched between the probe D2-branes and one of N D2-brane which involves the excitations of the gauge fields and scalar fields on both sides of the branes. 
The mass gap caused by the U(1) instanton is displayed in Fig. 3(a).
It is interesting to note that the mass gap of the U(1) gauge theory increases
as the mass of the matter field decreases while $g_{MY}$ is kept fixed.
This is contrary to the U(1) gauge theory coupled with `free' matter fields where lighter matters would be more effective in screening gauge field.
The opposite trend in the present case is the strong coupling effect of the additional SU(N) gauge field.
Even though mass of matter fields decreases, the increasing trend of the effective coupling in the 2+1D SU(N) gauge theory makes it harder for the matters to be polarized at lower energy.
This is an example showing that change in the dynamics of matter fields can drastically change their screening behavior.

\subsection{$p=6$} 
The parallel D6/D2 brane configuration preserve 8 supersymmetries\cite{CHERKIS,ERDMENGER}.  
This configuration is also stable because it is a BPS state.  
The 2+1D degrees of freedom consist of one vector multiplet, one neutral hyper multiplet
and $N$ fundamental hyper multiplets.
The scalars in the vector multiplet describes the transverse fluctuations of probe brane 
in the directions $x^7$, $x^8$ and $x^9$ and
the scalars in the neutral hypermultiplet, in the directions $x^3$, $x^4$, $x^5$ and $x^6$.
The $N$ fundamental hyper multiplets are stretched string modes.
The neutral hyper multiplet, and the fermions and the scalars in the vector multiplet are suppressed in the effective action Eq.(\ref{Gamma}).
The mass of the matter field is again given by $\Lambda$.  
The conditions for the small curvature and small string coupling (\ref{smallRg})
becomes 
\bq
\left( \frac{1}{g_{YM}^2 N} \right)^{1/3} <<  \Lambda  <<  \frac{N}{(g_{YM}^2)^{1/3}}.
\label{the_range6}
\eq
Note that the Yang-Mills coupling $g_{YM}^2$ has a dimension of $(length)^3$ in (6+1)-dimension.  
The lower bound of $\Lambda$ is the threshold between the strong coupling regime at high energy and the weak coupling regime at low energy.  
The U(1) gauge coupling and the inverse size of the instanton becomes
\bq
g^2  =  \frac{ 2 \Lambda }{ N },
\mbox{\hspace{0.5cm}}
M  =  \left( \frac{ 8 \pi^2 \Lambda}{g_{YM}^2 N } \right)^{\frac{1}{4}}.
\label{obs6}
\eq
The dimensionless gauge coupling $g^2 \Lambda^{-1}$ does not flow with $\Lambda$.  
Moreover the renormalized gauge coupling is independent of the Yang-Mills gauge coupling even in the strong coupling regime of the 6+1D Yang-Mills theory, that is, $g_{eff}^2 >> 1$.  
This is consistent with the one-loop result in the weak coupling regime $g^2 \Lambda^{-1} \sim 1/N$.
Thus it is likely that the dimensionless U(1) gauge coupling
does not flow in the whole range of energy scale including both the strong and
weak coupling regimes of the SU(N) gauge theory as is shown in Fig. 2(b).
If the U(1) gauge coupling has different value at high energy
it will be quickly renormalized to $g^2 \Lambda^{-1} \sim 1/N$ at low energy, which
is represented as dotted lines in Fig. 2(b).
The mass gap as a function of the normalized energy scale (mass of the matter fields) is displayed in Fig. 3(b).
At lower energies the 6+1 gauge coupling becomes weaker resulting in the smaller mass gap.
This is opposite to the $p=2$ case.

\subsection{$p=4$} 
The D4/D2-brane system breaks all
supersymmetries and the gravitational attraction renders this system unstable.
This can be seen from the $\Lambda$ dependence of the effective action in
Eq.(\ref{Gamma}) with $F$ set to be $0$.  
However here we are interested in the dynamics of the U(1) gauge field on the
D2-brane at a fixed position.
For this we suppress the transverse fluctuations of the D2-brane and treat $\Lambda$ as a
parameter.  
We will see a clue that the neglect of the transverse fluctuations in the gravity
description corresponds to the neglect of all unstable modes in the
full unstable field theory thus defining a well defined field theory problem.
If we ignore the tachyonic modes, $\Lambda$ can be
regarded as bare mass of the non-tachyonic matter fields which comes from the
stretching strings.  However in the non-supersymmetric case ($p=4$) the actual
mass of the matter field may be different from $\Lambda$ owing to the coupling
with the SU(N) gauge field.  This is especially true if the matter fields are
strongly coupled with the SU(N) gauge theory.

The conditions for the small curvature and small string coupling
(\ref{smallRg}) become
\bq
\frac{1}{g_{YM}^2 N} <<  \Lambda  << \frac{N^{\frac{1}{3}}}{g_{YM}^2}
\label{the_range1}
\eq
and the U(1) gauge coupling $g^2$ and the mass scale $M$ in the effective
action (\ref{Gamma}), 
\bq
g^2  =  \left( \frac{ g_{YM}^2 \Lambda^3 }{ 4 \pi^3 N } \right)^{\frac{1}{2}}, 
\mbox{\hspace{0.5cm}}
M  =  \left( \frac{\Lambda^3}{\pi g_{YM}^2 N } \right)^{\frac{1}{4}}.
\label{obs4}
\eq
The 4+1D gauge coupling has a dimension of length and becomes weaker as energy is lowered.  
The lower bound of $\Lambda$ in (\ref{the_range1}) is the threshold energy 
$\Lambda_c \sim \frac{1}{N g_{YM}^2}$ which divides the strong and weak coupling regimes of the 4+1D SU(N) gauge theory.
The gravity solution is valid only in the strong coupling regime ($\Lambda >> \Lambda_c$).  
In this region the dimensionless gauge coupling scales as 
$g^2 \Lambda^{-1}   \sim   \left( \frac{\Lambda g_{YM}^2}{N}  \right)^{\frac{1}{2}}$.
The flow of the U(1) gauge coupling is shown in Fig. 2(c).
The mass gap of the compact U(1) gauge theory is displayed in Fig. 3(c). 
As in the case of $p=6$ the mass gap decreases with decreasing energy scale.

It is instructive to compare to
the case where the 4+1D SU(N) gauge theory decouples and
$N$ species of matter fields with mass $\Lambda$ are coupled
only with the U(1) gauge field.  
In this case the one-loop effect renormalizes the U(1) gauge coupling to
$g^2 \sim \frac{\Lambda}{N}$\cite{IOFFE,MURTHY,WEN02,SENTHIL04,HERMELE} and the dimensionless gauge coupling at
the energy scale $\Lambda$ does not flow with the energy scale. 
Theories with different gauge couplings flow to the fixed point at low energy.
This is displayed in Fig. 4 which is essentially the same as the one in $p=6$ case (Fig. 2(b)).
The reason why the U(1) gauge coupling decreases with decreasing energy in the presence
of SU(N) gauge field can be explained in the following way.  At lower energy
the SU(N) gauge coupling becomes weaker.  As a result the matter fields become
more polarizable and more effective in screening the U(1) gauge field leading
to the decreasing behavior of the gauge coupling.
In the non-supersymmetric background ($p=4$) there is no cancellation between bosonic and fermionic fields.
In this case the SU(N) gauge field play an important role in determining the U(1) gauge coupling in the probe brane.
This is contrary to the supersymmetric $p=6$ case where there is no flow of dimensionless gauge coupling even in the strong coupling regime of the SU(N) gauge theory.
It is interesting to note that the gravity solution in (\ref{obs4}) predicts the U(1) gauge coupling at the threshold energy $\Lambda_c$ to be $g^2 \sim \frac{\Lambda}{N}$ which is consistent with the prediction of the weakly coupled SU(N) gauge theory.
This is a nontrivial consistent check to our earlier assumption
that the gravity solution with the neglect of the unstable mode
describes the 2+1D/4+1D U(1)/SU(N) theory with fundamental matters
in the strong coupling regime of the SU(N) theory.
Even though the gravity solution begins to loose its validity around the
threshold the qualitative feature is captured.

It is reminded that $\Lambda$ is not necessarily the same as the mass of the
matter field in the strong coupling regime because there is no supersymmetry for the D2/D4 case.
Therefore it is hard to directly
interpret the scaling dimension of $g^2$ and $M$ in Eq.(\ref{obs4}).  
However the ratio between the scaling dimension is meaningful,
\bq
\frac{d \ln(M \Lambda^{-1})}{d \ln(g^2 \Lambda^{-1})} = -\frac{1}{2}
\eq
because the ratio is independent of definition of $\Lambda$.  This exponent
$-1/2$ shows how the mass scale associated with the instanton size scales
relative to the gauge coupling.

\section{conclusion}
In summary, we studied how a change in the dynamics of fundamental matter fields
caused by strong coupling to SU(N) gauge field changes their screening property
in the 2+1D compact U(1) gauge theory.
For this, we considered the probe action of a D2-brane in the gravity
background dual to a large number of coincident Dp-branes by treating the
separation between the branes as a parameter.  
We studied the effects of the
SU(N) gauge field of the Dp-branes on the dynamics of the 2+1D compact compact
U(1) gauge field of the D2-brane as the effective coupling strength of the
SU(N) gauge theory is tuned by the separation.  We determined the gauge
coupling, the size of instanton and the mass gap of the 
non-supersymmetric compact U(1) gauge theory as a function of the separation.  
The results are interpreted in terms of the 2+1D U(1) gauge theory and the
p+1D SU(N) gauge theory which are coupled with each other through
a large number of matter fields in fundamental representation of both U(1) and
SU(N) gauge groups.  
It is found that the strong coupling of the matter fields to the SU(N) gauge field
can drastically modify the dynamics of the U(1) gauge field.
In the supersymmetric D6/D2 brane system 
the renormalized U(1) gauge coupling is shown to be 
independent of the (6+1)-dimensional
Yang-Mills coupling even in the strong coupling regime.
For D4/D2 case it is shown that the dimensionless
U(1) gauge coupling decreases with decreasing separation in the strong
coupling regime for the SU(N) gauge theory and that it is continuously
connected with the value in the weak coupling regime.

\section*{ACKNOWLEDGEMENTS}
We would like to thank P. A. Lee, A. Hanany and T. Senthil for helpful discussions.  
We are especially grateful to H. Liu for various advices at an early stage of this work.  
S.L. was supported by the Post-doctoral
Fellowship Program of Korea Science $\&$ Engineering Foundation (KOSEF) and
NSF grant DMR-0201069.
X.G.W. was supported by NSF Grant No. DMR--01--23156,
NSF-MRSEC Grant No. DMR--02--13282, and NFSC No. 10228408.


\newpage
\centerline{FIGURE CAPTIONS} 
\begin{itemize}

\item[Fig. 1]
Examples of (a) a disk diagram and (b) a cylinder diagram in the loop expansion of string theory (left column) which correspond to a planar and a non-planar diagrams respectively in the field theory (right column).
In the left column, the plane represents the probe D2-brane.
The half sphere and the cylinder represents the string world sheet.
$F_{\mu \nu}$ denotes the background U(1) gauge field on the D2-brane and $G_{\mu \nu}$, the background metric generated by N Dp-branes.
In the right column, the double line with two solid lines represents the propagator of U(1) gauge field,
the double line with one solid and one dashed line, that of the fundamental matter fields and
the double line with two dashed lines, that of the SU(N) gauge fields.

\item[Fig. 2 ]
Flow of the dimensionless U(1) gauge coupling as a function of the energy scale in the 2+1D/p+1D U(1)/SU(N) gauge theory for (a) $p=2$, (b) $p=6$ and (c) $p=4$.
$g_{eff}^2 = g_{YM}^2 N \Lambda^{p-3}$ is the effective gauge coupling for the p+1D SU(N) gauge theory, 
${\cal R} l_s^2$, the dimensionless scalar curvature of the metric dual to the Dp-branes and
$e^\phi$, the local string coupling. 
The solid line denotes the flow of the U(1) gauge coupling for the 2+1D/p+1D U(1)/SU(N) gauge theory
realized by the D2/Dp-brane system.
The dotted line denotes the flow of the U(1) gauge coupling for general 2+1D/p+1D U(1)/SU(N) gauge theory
which initially has different U(1) gauge coupling at high energy.

\item[Fig. 3 ]
The mass gap of the U(1) gauge field in the IIA gravity regime as a function of the dimensionless energy scale $\lambda = \Lambda (g_{YM}^2)^\frac{1}{(p-3)} N^{\frac{1}{p-7}}$ for (a) $p=2$, (b) $p=6$ and (c) $p=4$.
( Note that the mass gap is plotted as a function of the inverse of the normalized energy scale for $p=2$ in order to fit the IIA gravity regime within the interval from $0$ to $1$.)
Logarithmic scale is used for the vertical axis.

\item[Fig. 4 ]
Flow of the U(1) gauge coupling in the 2+1D U(1) gauge theory coupled with matter field without 
the additional SU(N) theory.
The solid line denotes the gauge coupling at the conformal fixed point and
the dotted line, the flow of the U(1) coupling which initially has different value from the fixed point value.

\end{itemize}

\newpage

\begin{figure}
        \includegraphics[height=10cm,width=15cm]{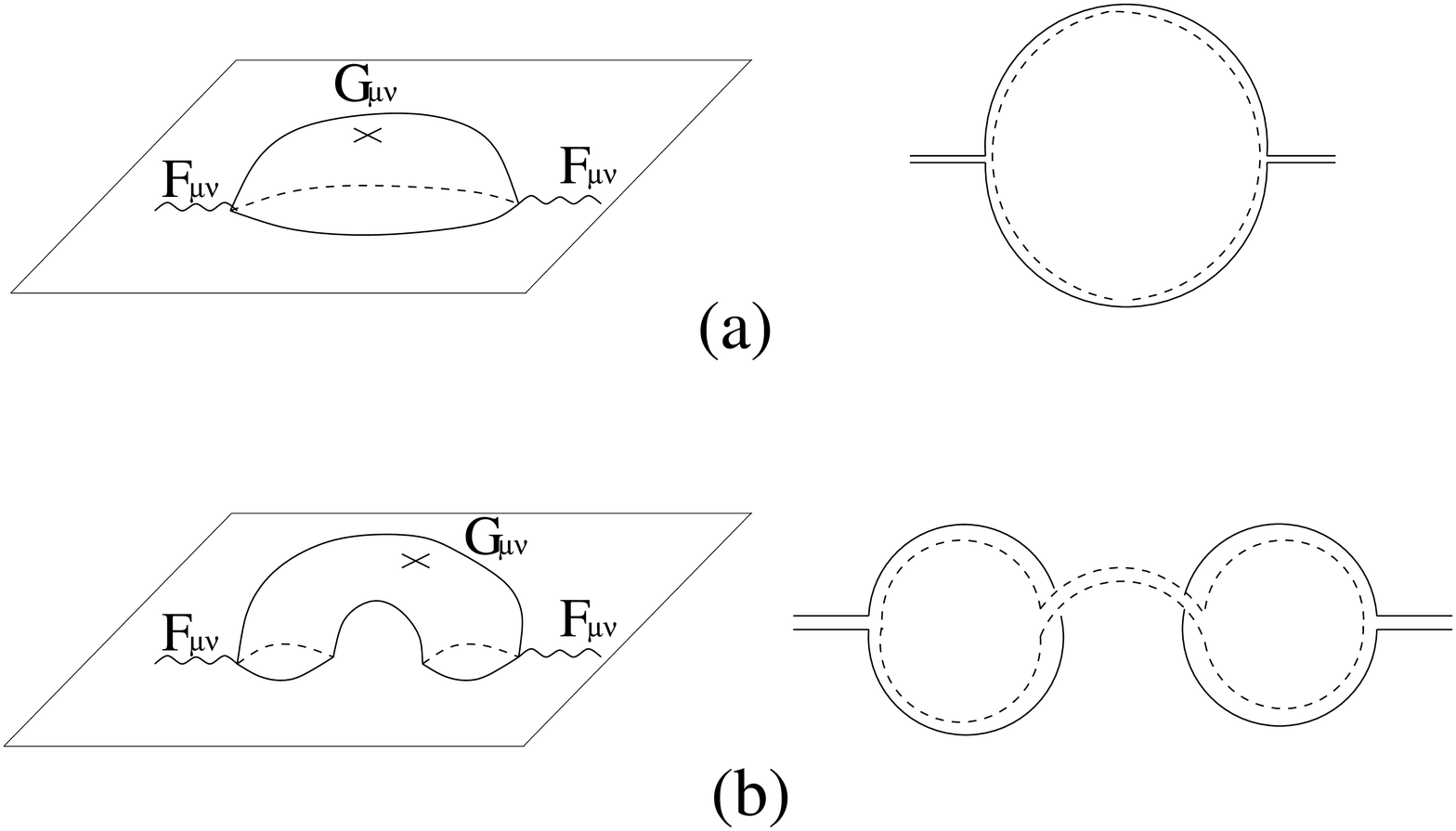}
\label{fig:1}
\caption{}
\end{figure}

\begin{figure}
        \includegraphics[height=7.5cm,width=12cm]{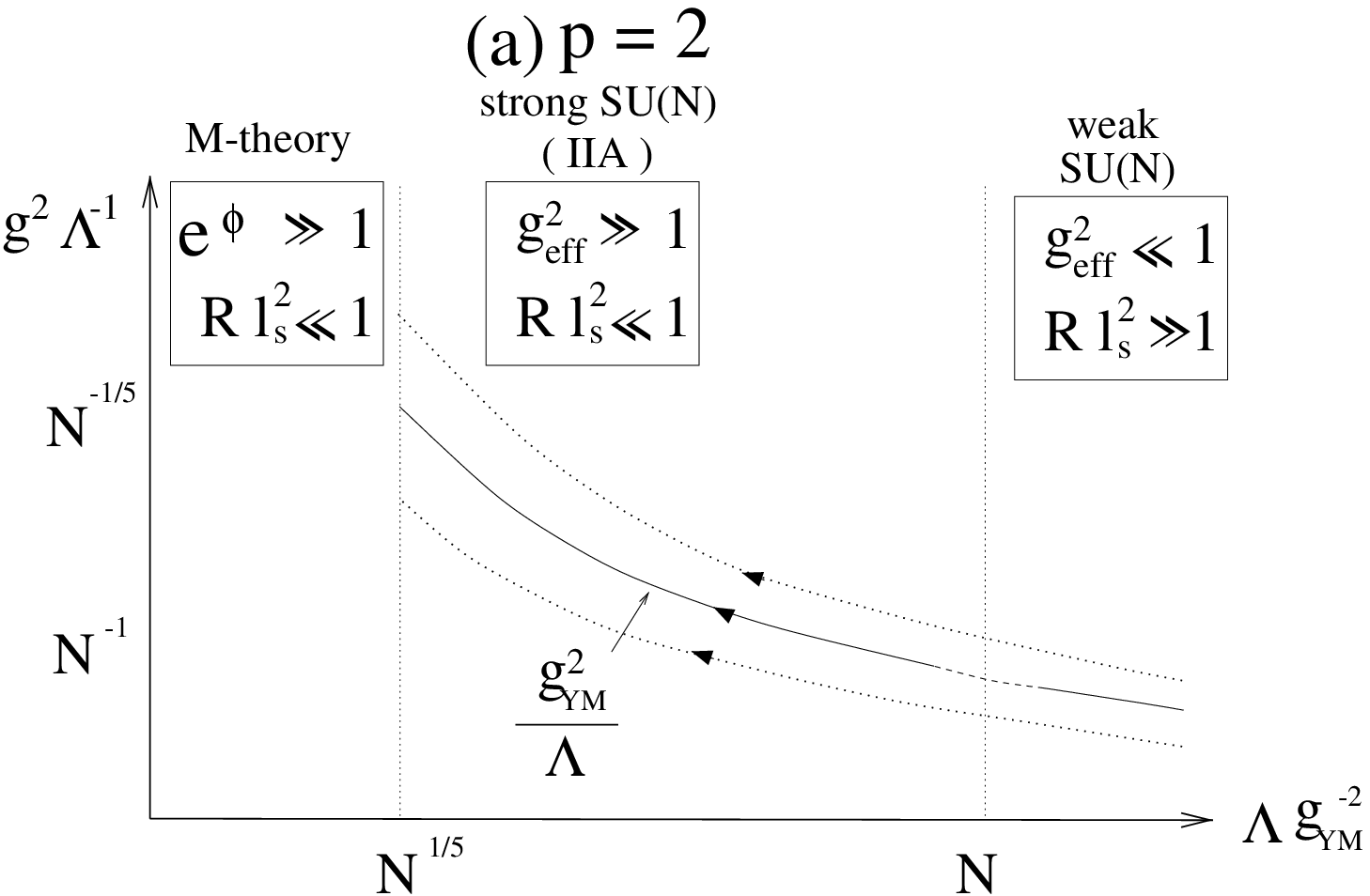}
        \includegraphics[height=7.5cm,width=12cm]{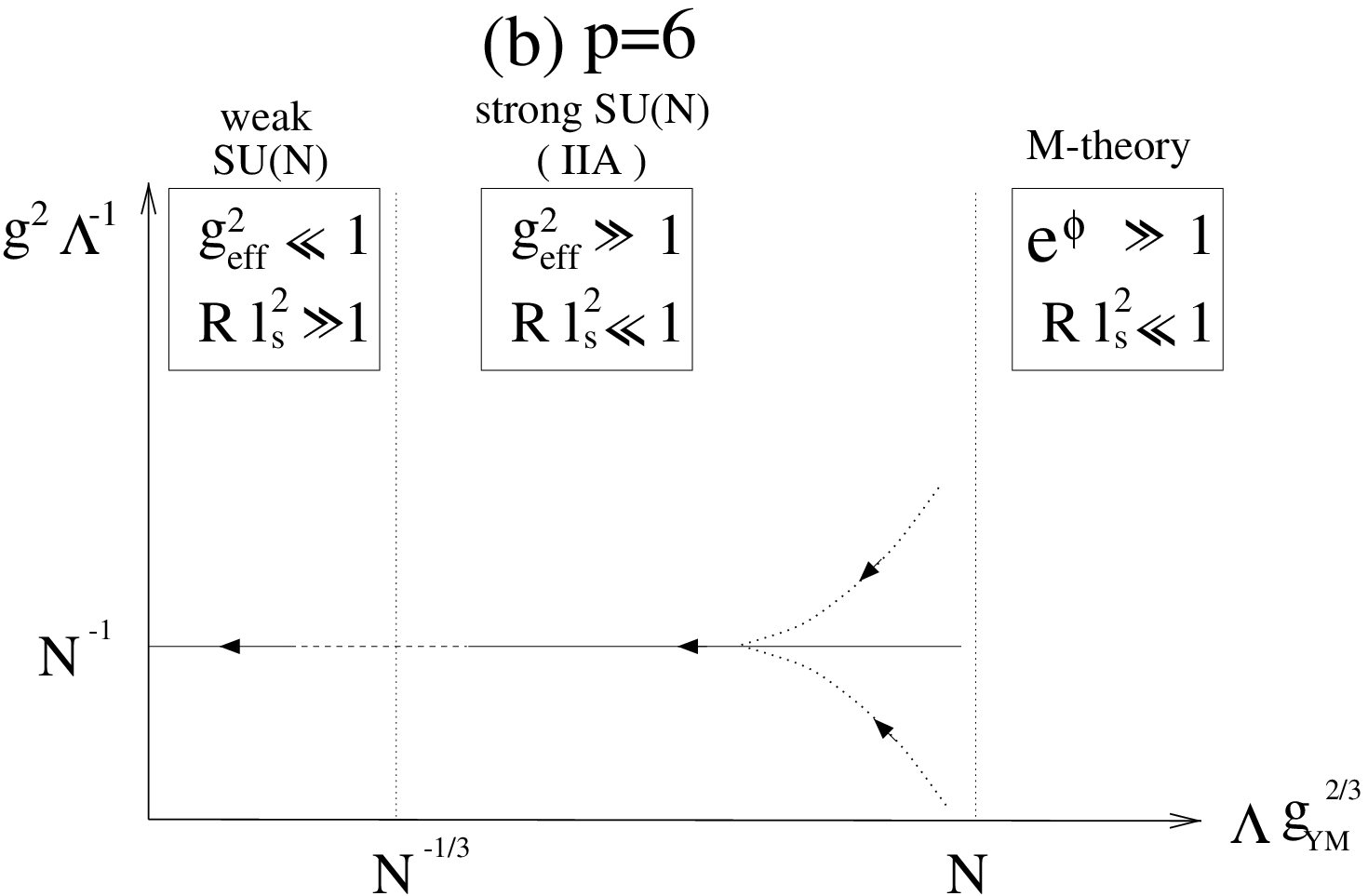}
        \includegraphics[height=7.5cm,width=12cm]{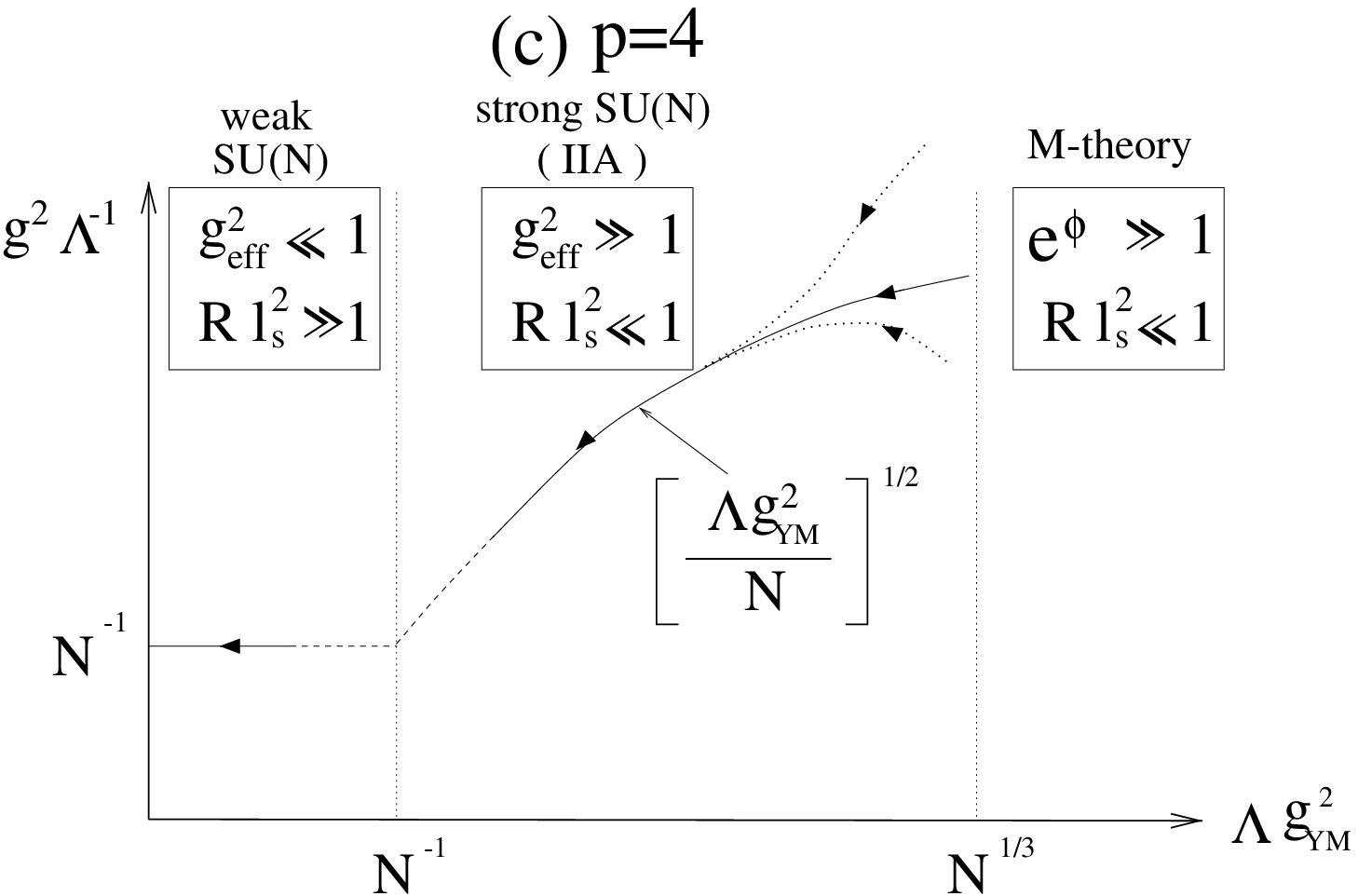}
\label{fig:2}
\caption{}     
\end{figure}

\begin{figure}
        \includegraphics[height=7.5cm,width=12cm]{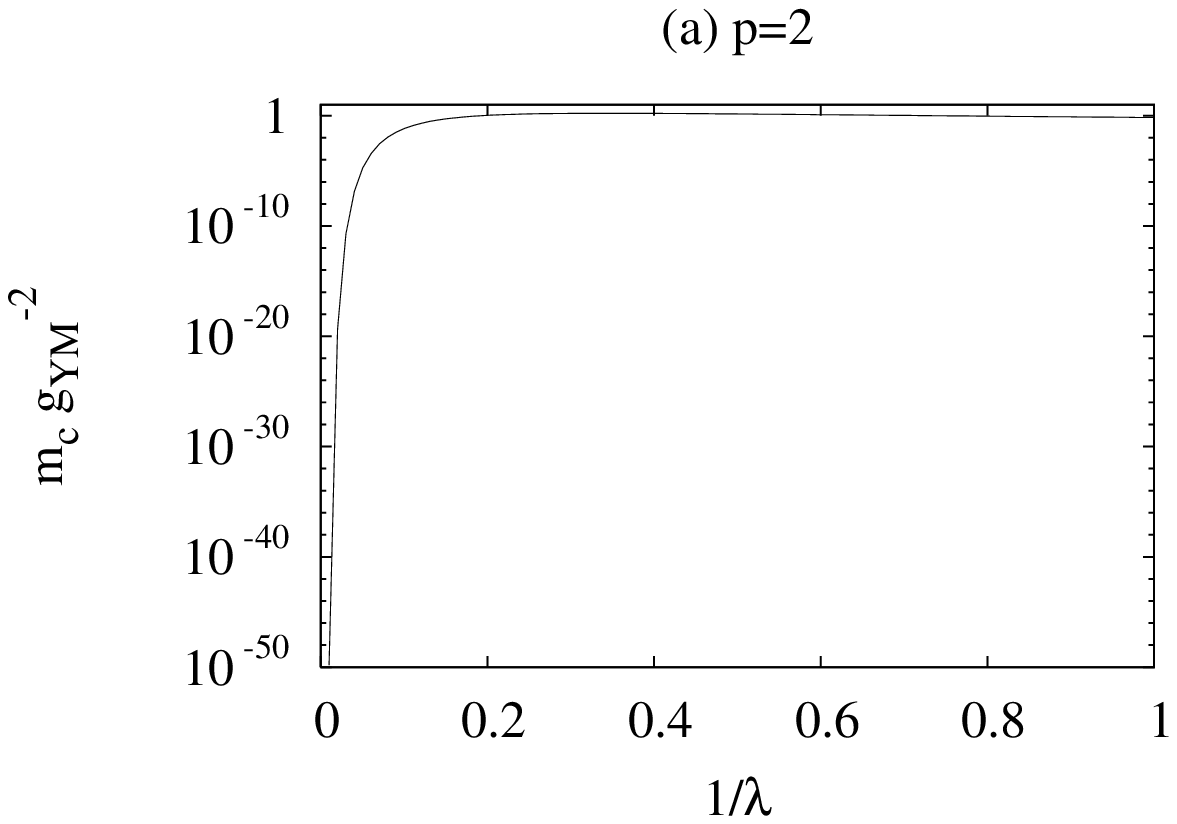}
        \includegraphics[height=7.5cm,width=12cm]{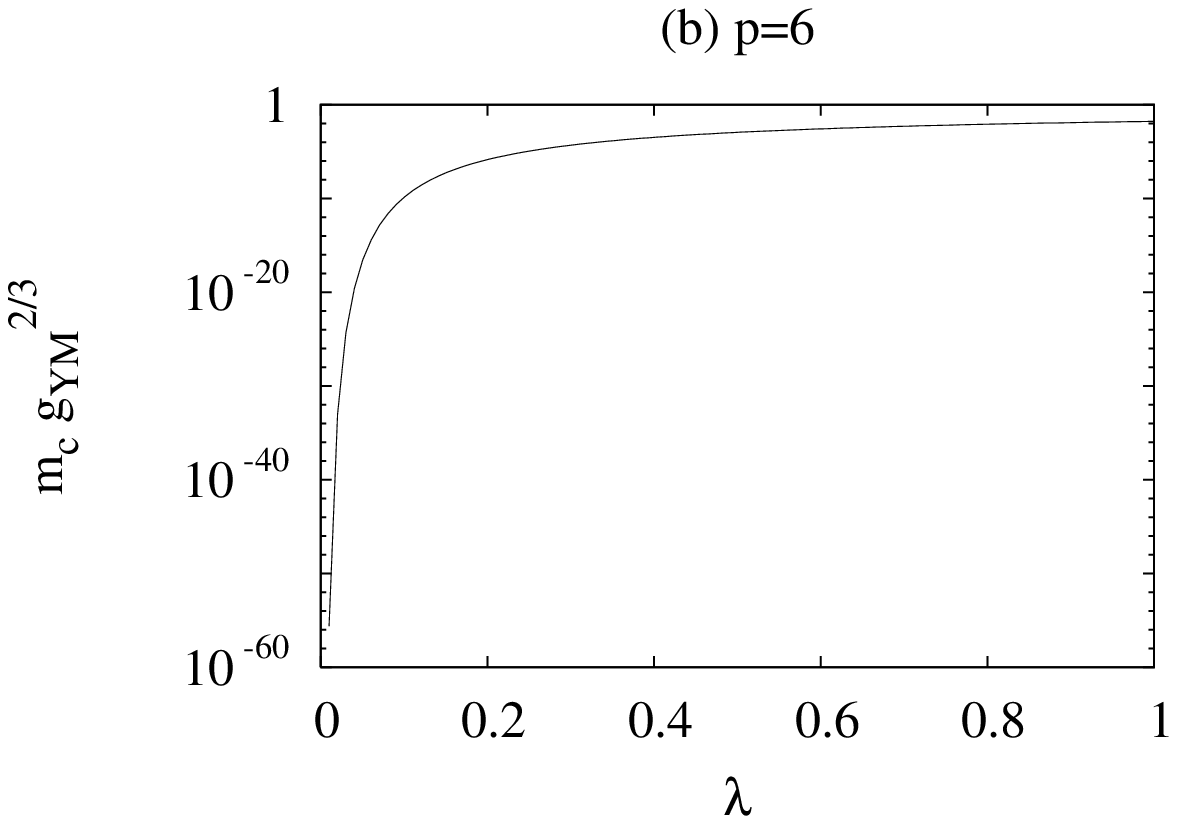}
        \includegraphics[height=7.5cm,width=12cm]{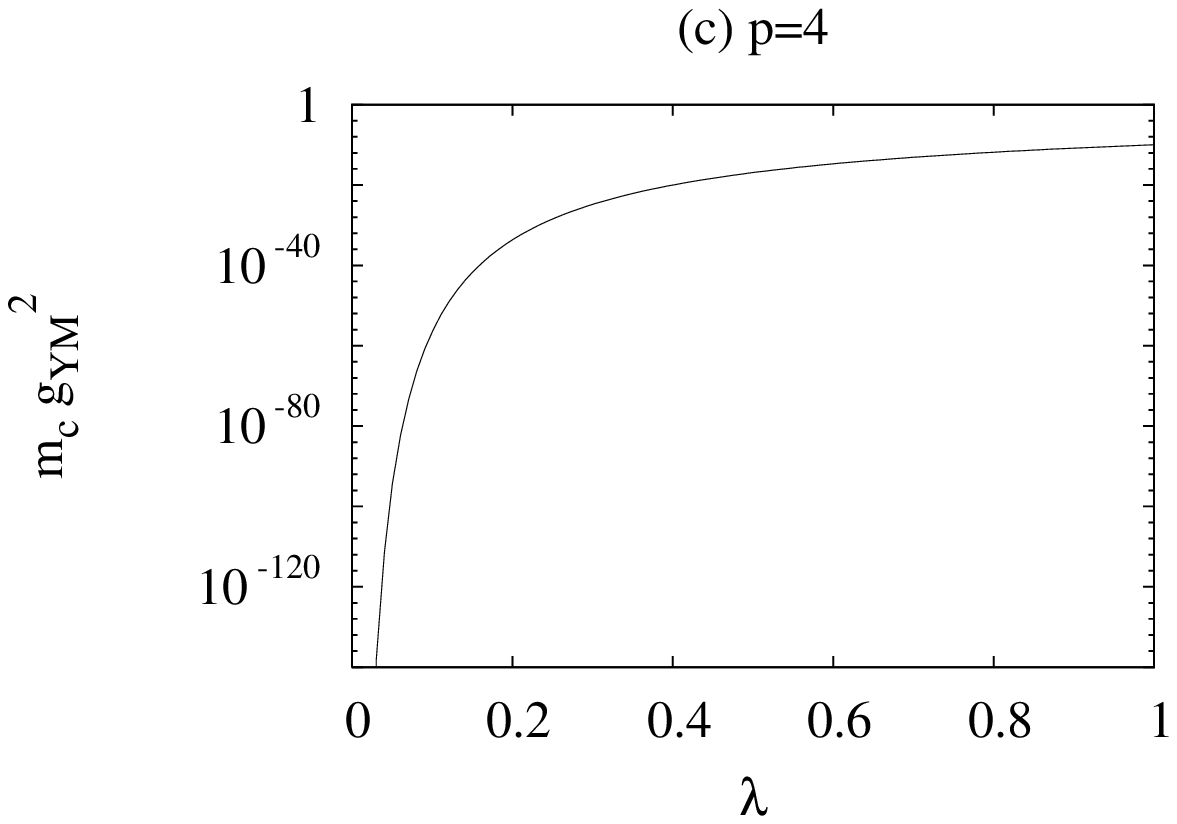}
\label{fig:3}
\caption{}
\end{figure}

\begin{figure}
        \includegraphics{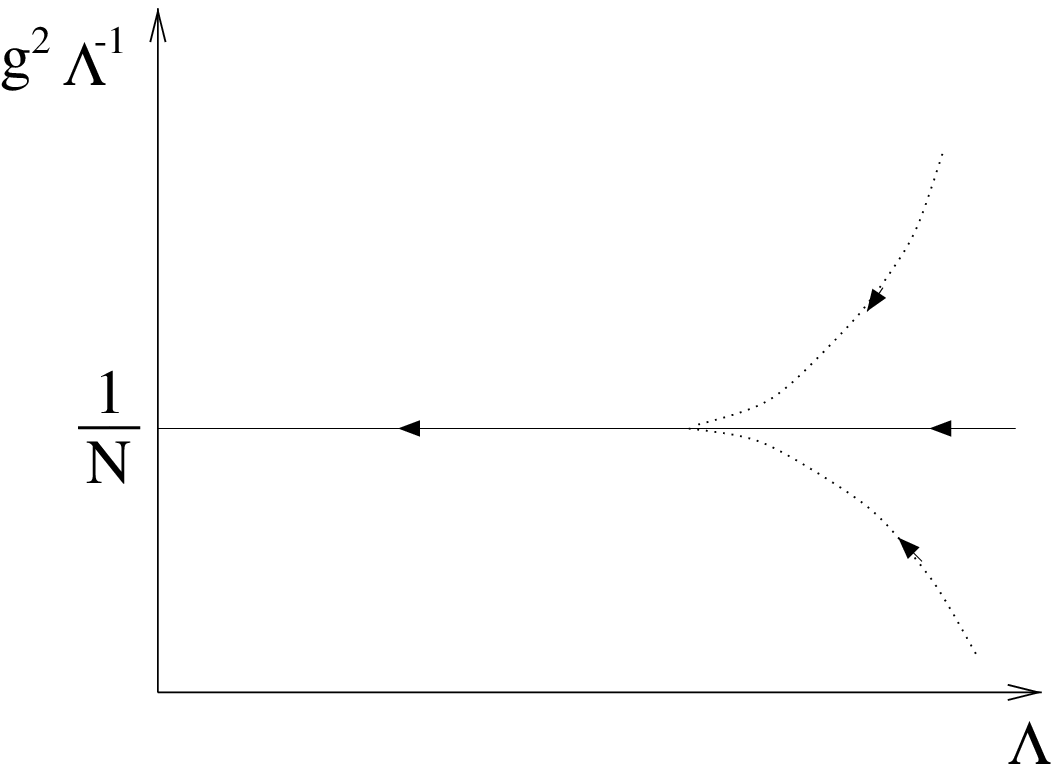}
\label{fig:4}
\caption{}     
\end{figure} 


\begin{thebibliography}{27}


\bibitem{POLYAKOV77} A. M. Polyakov, Phys. Lett. B 59 (1975) 82; Nucl.  Phys. B 120 (1977) 429. 
\bibitem{EINHORN} M. B. Einhorn and R. Savit, Phys. Rev. D 19 (1979) 1198.
\bibitem{FRADKIN79} E. Fradkin and S. H. Shenker, Phys. Rev. D 19 (1979) 3682.
\bibitem{NAGAOSA93} N. Nagaosa, Phys. Rev. Lett. 71 (1993) 4210. 
\bibitem{MUDRY} C. Mudry and E. Fradkin, Phys. Rev. B 49 (1994) 5200; Rev. Rev. B 50 (1994) 11409.
\bibitem{NAYAK} C. Nayak, Phys. Rev. Lett. 85 (2000) 178.
\bibitem{NAGAOSA00} N. Nagaosa and P. A. Lee, Phys. Rev. B 61 (2000) 9166.
\bibitem{ICHINOSE} I. Ichinose and T. Matsui, Phys. Rev. Lett. 86 (2001) 942.
\bibitem{WEN02} X.-G. Wen, Rev. Rev. B 65 (2002) 165113; references there-in. 
\bibitem{RWspin} W. Rantner and X.-G. Wen, Phys. Rev. B 66 (2002) 144501.
\bibitem{HERBUT} I. F. Herbut and B. H. Saradjeh, Phys. Rev. Lett. 91 (2003) 171601; 
I. F. Herbut, B. H. Saradjeh, S. Sachdev and G. Murthy, Phys.  Rev. B 68 (2003) 195110. 
\bibitem{KLEINERT} H. Kleinert, F. S. Nogueira and A. Sudbo, Nucl. Phys. B 666 (2003) 361.
\bibitem{HERMELE} M. Hermele, T. Senthil, M. P. A. Fisher, P. A. Lee, N. Nagaosa and X.-G. Wen, Phys. Rev. B 70 (2004) 214437.
\bibitem{SENTHIL04} T. Senthil, A. Vishwanath, L. Balents, S. Sachdev and M. P. A. Fisher, Science 303 (2004) 1490.
\bibitem{IOFFE} L. B. Ioffe and A. I. Larkin, Phys. Rev. B 39 (1989) 8988.
\bibitem{MURTHY} G. Murthy and S. Sachdev, Nucl. Phys. B 344 (1990) 557.
\bibitem{GHOSH} S. Ghosh, J. Phys. A: Math. Gen. 33 (2000) 1915.
\bibitem{ABREU} E. M. C. Abreu, D. Dalmazi, A. S. Dutra and M. Hott, Phys.  Rev. D 65 (2002) 125030.
\bibitem{AHARONY} O. Aharony, S. S. Gubser, J. Maldacena, H. Ooguri and Y. Oz, Phys. Rep. 323 (2000) 183.
\bibitem{POLYAKOV98} A. M. Polyakov, Nucl. Phys. B Proc. Suppl. 68 (1998) 1.
\bibitem{MALDACENA} J. M. Maldacena, Adv. Theor. Math. Phys. 2 (1998) 231.
\bibitem{GUBSER} S. S. Gubser, I. R. Klebanov and A. M. Polyakov, Phys. Lett. B 428 (1998) 105.
\bibitem{WITTEN} E. Witten, Adv. Theor. Math. Phys. 2 (1998) 253.
\bibitem{WITTEN2} E. Witten, Adv. Theor. Math. Phys. 2 (1998) 505.
\bibitem{KARCH} A. Karch and E. Katz, J. High Energy Phys. 06 (2003) 053.
\bibitem{BABINGTON} J. Babinton, J. Erdmenger, N. Evans, Z. Guralnik and I. Kirsch, Phys. Rev. D 69 (2004) 066007.
\bibitem{CHERKIS} S. A. Cherkis and A. Hashimoto, J. High Energy Phys. 11 (2002) 036.
\bibitem{NUNEZ} C. Nunez, A. Paredes and A. V. Ramalo, hep-th/0311201.
\bibitem{ERDMENGER} J. Erdmenger and I. Kirsch, hep-th/0408113.
\bibitem{KRUCZENSKI} M. Kruczenski, D. Mateos, R. C. Myers and D. J. Winters, J. High Energy Phys. 05 (2004) 041.
\bibitem{BENVENUTI} S. Benvenuti, S. Franco, A. Hanany, D. Martelli and J. Sparks, hep-th/0411264.
\bibitem{SEIBERG} N. Seiberg, Nucl. Phys. B (Proc. Suppl.) 67 (1998) 158; references there-in. 
\bibitem{ASEN} A. Sen, J. High Energy Phys. 08 (1998) 012; J. High Energy Phys. 12 (1999) 027; K. Takahashi, hep-th/0404205.
\bibitem{THOOFT} G. 't Hooft, Nucl. Phys. B 72 (1974) 461.
\bibitem{ITZHAKI} N. Itzhaki, J. M. Maldacena, J. Sonnenschein and S. Yankielowicz, hep-th/9802042.
\bibitem{DBI} E. S. Fradkin and A. A. Tseytlin, Phys. Lett. B 163  (1985) 123; Phys. Lett. B 158 (1985) 316; Nucl. Phys. B 261 (1985) 1; C. G. Callan, C. Lovelace, C. R. Nappi and S. A. Yost, Nucl. Phys. B 308 (1988) 221.
\bibitem{GOPFERT} M. Gopfert and C. Mack, Commun. Math. Phys. 82 (1982) 545.
\bibitem{INSTANTON} J. Polchinski and P. Pouliot, Phys. Rev. D 56 (1997) 6601; 
N. Dorey, V. V. Khoze and M. P. Mattis, Nucl. Phys. B 502 (1997) 94;
M. Dine and N. Seiberg, Phys. Lett. B 409 (1997) 239.

\end{thebibliography}
\end{document}